\documentclass[traditabstract,twocolumn]{aa}
\usepackage{graphicx}
\usepackage{longtable}
\usepackage{lscape}
\usepackage{rotating}
\usepackage{enumerate}
\usepackage{float}
\usepackage{placeins}
\usepackage{amsmath}
\usepackage{color}
\usepackage{footnote}
\usepackage{pdflscape}

\def\lesssim{\mathrel{\hbox{\rlap{\hbox{\lower4pt\hbox{$\sim$}}}\hbox{$<$}}}}
\def\gtrsim{\mathrel{\hbox{\rlap{\hbox{\lower4pt\hbox{$\sim$}}}\hbox{$>$}}}}
\begin{document}
   \title{The physics and kinematics of the  evolved, interacting \\ planetary nebula PN G342.0-01.7}

\author{Alaa Ali\inst{1,2}\and Morsi A. Amer\inst{1,2}\and M.A. Dopita\inst{3,1} \and F.P.A. Vogt\inst{3} \and H. M. Basurah\inst{2}}
\institute{Astronomy Dept, Faculty of Science, King Abdulaziz University, Jeddah, Saudi Arabia \label{inst1}
\and Department of Astronomy, Faculty of Science, Cairo University, Egypt \label{inst2}
\and Research School of Astronomy and Astrophysics, Australian National University, Cotter Rd., Weston ACT 2611, Australia\label{inst3}}

\offprints{Alaa Ali}
  \date{Received  2015; accepted 00, 00}

 \abstract {Integral field spectroscopy has been obtained for very few evolved planetary nebulae (PNe). Here we aim to study the physical and kinematical characteristics of the unstudied old planetary nebula (PN) PN G342.0-01.7, which shows evidence of interaction with its surrounding interstellar medium. We used Integral Field Spectra from the Wide Field Spectrograph on the ANU 2.3 m telescope to provide spectroscopy across the whole object covering the spectral range 3400-7000 {\AA}. We formed narrow-band images to investigate the excitation structure. The spectral analysis shows that the object is a distant Peimbert  Type I PN of low excitation, formally of excitation class of 0.5. The low electron density, high dynamical age, and low surface brightness of the object confirm that it is observed fairly late in its evolution. It shows clear evidence for dredge-up of CN-processed material  characteristic of its class. In addition, the  low peculiar velocity of 7 km s$^{-1}$ shows it to be a member of the young disk component of our Galaxy. We further determined an average expansion velocity of $V_{exp} = 20.2\pm 1.3$ km s$^{-1}$, a local standard of rest radial velocity $RV_{LSR} = -27.7\pm1.7$ km s$^{-1}$, and  a distance of $2.06\pm 0.6$ kpc for the object. We built a self-consistent photoionisation model for the PNe matching the observed  spectrum, the H$\beta$ luminosity, and the diameter. On the basis of this we derive an effective temperature $\log T_{\rm eff} \sim 5.05$ and luminosity $1.85 < \log L < 2.25$. The temperature is much higher than might have been expected using the excitation class, proving that this can be misleading in classifying evolved PNe. PN G342.0-01.7 is in interaction with its surrounding interstellar medium through which the object is moving in the south-west direction. This interaction drives a slow shock into the outer PN ejecta. A shock model suggests that it only accounts for about 10\% of the total luminosity, but has an important effect on the global spectrum of the PN.


}

\authorrunning{Ali et al.}
\titlerunning{The planetary nebula PN G342.0-01.7}
  \maketitle
%

\section{Introduction}
The planetary nebula PN G342.0-01.7 was discovered and classified as a true planetary nebula in the Macquarie/AAO/Strasbourg H$\alpha$ Planetary Nebula Catalogue (MASH I) where it was given the name PHR1702-4443 (Parker et al. 2006). Over 1200 true, likely and possible new Galactic planetary nebulae were discovered in this AAO/UKST H$\alpha$ survey of the southern Galactic plane. The MASH I \& MASH II (Miszalski et al., 2008) catalogues increase the number of known Galactic PNe by $\sim 60\%$, which represents the largest ever incremental sample. According to the morphological scheme used in the MASH catalogue, the object was classified as an elliptical PN with apparent angular size 90.0\arcsec x 63.0\arcsec. Furthermore, it is subclassified as a one-sided enhancement/asymmetry with a well defined ``Ear-like'' ring structure.  The object shows a broken shell with extended diffuse emission seen across a highly filamentary shell structure (Parker et al. 2006). Weidmann \& Diaz (2008) determined the spatial orientation of the object within the Galaxy. They measured the position angle $PA$ of the projected major axis over the sky with respect to the equatorial coordinate system $EPA = 90^{\circ}$. The EPA was transformed into the galactic system $GPA = 142^{\circ}$. This object represents a good example of a low surface brightness PN caught at a late stage of its evolution, and possibly in an early stage of interaction with the surrounding ISM. Due to their extreme faintness and large spatial extent, such objects have not been spectroscopically studied in great detail hitherto, although Danehkar, Parker \& Erolano (2013) have examined one such object, SuWt 2, in detail.

Studying evolved PNe and their central stars provides valuable constraints on the evolution from the PN towards the White Dwarf (WD) phase. Whenever these objects show signs of interaction with the ISM, they become important in understanding how the nuclear processed materials ejected in the PNe are returned to the ISM (Tweedy et al. 1995, Tweedy \& Kwitter (1996);  Xilouris et al. 1996; Kerber et al. 1998; Guerrero et al. (1998); Kerber et al. 2000; Ali et al. 2012).

The composition of the nuclear processed materials ejected into the ISM in the Asymptotic Giant Branch (AGB) phase and ionised in the PNe phase are strongly dependent on both the initial abundances in the precursor star and on the precursor mass, most recently discussed by Karakas, 2014 (see also references therein to earlier work). In particular, in the more massive, younger population of PNe, both Nitrogen and Helium are strongly enhanced by the operation of hot-bottom burning during helium shell flashes with the subsequent dredge-up of these elements into the outer envelope. The PNe with these more massive precursors are distinct both kinematically and morphologically from other PNe, as first recognised by Greig (1971). Their morphologies are characterized by the appearance of filamentary structure. They also frequently display a bi-polar structure. This may have a number of origins. These include rotation in the precursor star (e.g. Georgiev et al. 2011), binarity of the central star and interaction of their stellar winds (Dijkstra, \& Speck, 2006; Riera et al. 2012), and/or magnetic shaping (Garcia et al.1999). Kinematically they belong to the thin-disk population, as is to be expected from the relatively short lifetimes of the precursor stars. They show a small deviation between the velocity determined from the observed local standard of rest  radial velocity and that determined from the rotation curve (Greig 1972, Peimbert and Torres-Peimbert 1983). The chemical characteristics of this class of PNe was first recognised by Peimbert (1978) and they are now known as the Peimbert Type I PNe.

The evolved Type I PNe, although rather numerous, have been very poorly studied. Recently Frew et al (2006) discussed two nearby planetary nebulae, similar to the PN G342.0-01.7 in the sense they were classified as Type I PNe and show low surface brightness, electron density,  excitation class, and peculiar velocities. These have the classical bipolar morphologies and high central temperatures, which probably places them in the intermediate age category $(\sim 20000$~yr). Manteiga et al. (2004) studied another two evolved PNe of low surface brightness towards the Galactic bulge. Both show high N/O ratios and He abundances, similar to those observed in Type I disk PNe. A large low surface brightness annular PN was detected by Pierce et al. (2004), which shows a very low density and excitation class with early stage of interaction with the ISM.

The main objective of the present study  is to investigate the global physical and kinematical characteristics of the PN G342.0-01.7. For this we have made a mosaic of observations covering the whole nebula using  the Wide Field Spectrograph (WiFeS) - an integral field spectrograph mounted on the 2.3m telescope at Siding Spring Observatory. The major advantage of such observations is that one can extract a global spectrum of the entire nebula, avoid contamination by of the nebular emission by stars, and derive the global physical and kinematical nebular properties such as electron temperature, density, chemical abundances, radial and expansion velocities. In addition, it is easy to construct an emission line maps for the object in chosen emission lines such as [O III], [N II], [S II], $H\alpha$, and $H\beta$.

In this paper, we present our observational data in Section \ref{Obs}, discuss the general global properties of PN G342.0-01.7 and evidences for its interaction with its local ISM in Section \ref{Props}, and in Section \ref{Model} we present the results of detailed photoionisation and shock modelling of the ejecta. Finally in Section \ref{Conc} we present our conclusions.


\section{Observations \& data reduction}\label{Obs}
\subsection{ESO observations}
\begin{figure*}[htb!]
\centerline{\includegraphics[scale=0.5]{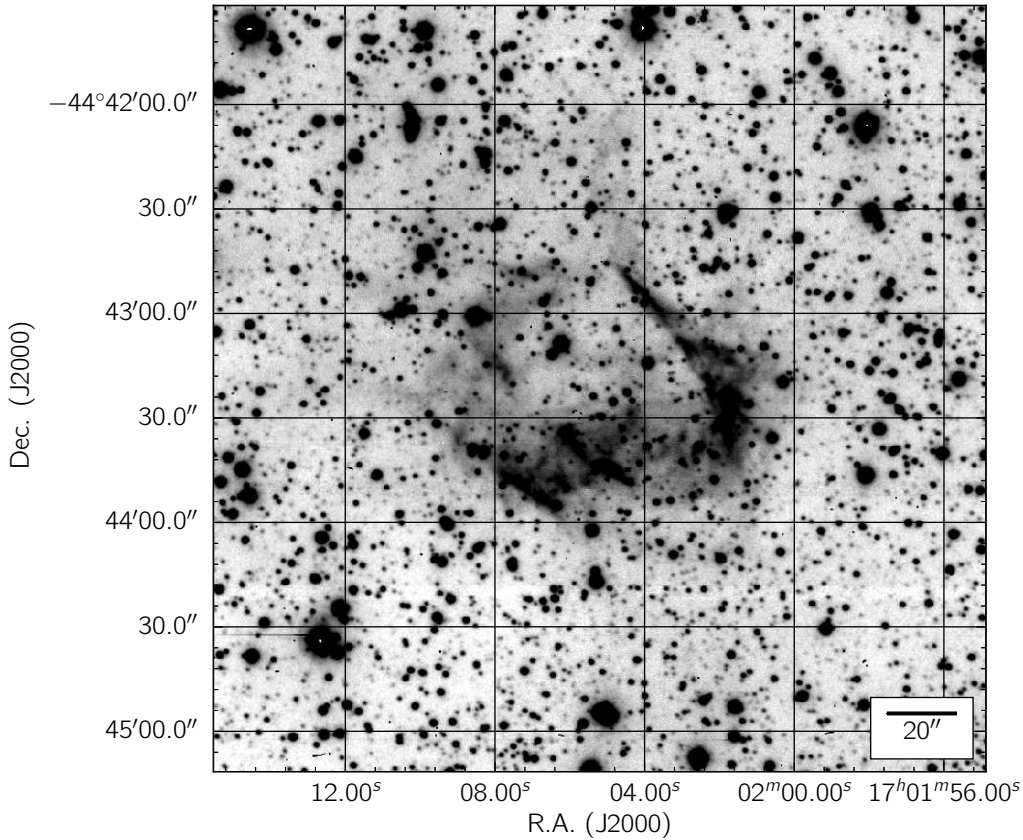}}\label{fig:FORS2}
\caption{H$\alpha$+[N\,{\sc\small II}] emission from PN G342.0-01.7 acquired with FORS2 on the VLT. The brightest emission clumps are oriented along a North-East to South-West axis, with additional diffuse emission forming a shell-like structure. Faint whiskers are also extending to the North and North-East from the brightest features.}\label{fig:FORS2}
\end{figure*}
 The best overall image of PN G342.0-01.7 has been acquired with the FORS2 instrument (Appenzeller et al. 1998) mounted on the Cassegrain focus of the Unit 1 (Antu) of the Very Large Telescope in Chile. PN G342.0-01.7 was observed with FORS2 in Period 89 (P.Id.: 089.D-0357(A); P.I.: Boffin) as part of series of observations of PNe in the MASH catalogue (Parker et al. 2006).

The extended structure of PN G342.0-01.7 is revealed in the H$\alpha$+[N\,{\sc\small II}] image (Figure~\ref{fig:FORS2}) which is comprised of two single 150s exposures acquired on 2012 July 11 with the ``H$\alpha$+83'' narrow-band filter ($\lambda_0=6563$\,\AA, $\lambda_{FWHM}=61$\,\AA) and downloaded from the ESO archive.  We reduced both exposures, offset along the North-South direction by 1\,arcmin to cover the gap between the two FORS2 CCDs in imaging mode, using the dedicated FORS pipeline (v5.0.11) inside the ESO \textsc{reflex} environment (Freudling et al. 2013). Each four individual CCD images (two per exposure) were processed individually and later combined into one global, averaged mosaic using a custom \textsc{python} script. The data was acquired with $2\times2$ binning, so that pixels in the final mosaic have a size of $0.5\times0.5$ square arcsec, while the seeing at the time of the observation was of the order of $\sim$0.8 arcsec.

As noted above, the effective bandpass of this ESO image is $\lambda_{FWHM}=61$\,\AA. Thus, all of the [N\,{\sc\small II}] emission is included in the bandpass of the filter along with the H$\alpha$. Since the [N\,{\sc\small II}] emission is dominant in this object, the morphology of the PNe in the ESO image is dominated by [N\,{\sc\small II}].

Figure~\ref{fig:FORS2} shows that PN G342.0-01.7 is composed of a series of bright filaments oriented along a North-East to South-West direction, as well as fainter and shell-like diffuse emission. In particular, we note the presence of diffuse emission or an outer envelope - not observed by WiFeS - extending towards the North and North-East directions (at and around R.A.: 17$^h$2$^m$5.0$^s$; Dec.: -44$^{\circ}$42$^{\prime}$15.0$^{\prime\prime}$) with respect to the brightest clumps.

\subsection{WiFeS observations}
The integral field spectra of the PN G342.0-01.7 were obtained over the two nights of March 30-31, 2013 with the Wide Field Spectrograph (WiFeS; Dopita et al. 2007, 2010) mounted on the 2.3-m ANU telescope at Siding Spring Observatory. The WiFeS instrument delivers a field of view of 25\arcsec x 38\arcsec at a spatial sampling of either 1.0\arcsec x 0.5\arcsec or 1.0\arcsec x 1.0\arcsec, depending on the binning on the CCD. For these observations we used the 1.0\arcsec x 1.0\arcsec option. The observations covered the blue spectral range 3400-5700 {\AA}  with a low spectral resolution of $B \sim 3000$ that corresponds to a full width at half maximum (FWHM) of $\sim 100$ km/s, while in the red spectral range 5700-7000 {\AA} we used a higher spectral resolution $R \sim 7000$ corresponding to a FWHM of $\sim 42$ km/s. The wavelength scale was calibrated using the Cu-Ar arc Lamp with 40s exposure at the beginning and throughout the night, while flux calibration was performed using the STIS spectrophotometric standard stars HD 111980, HD 031128 and HD160617 \footnote{Available at : \newline {\tt www.mso.anu.edu.au/~bessell/FTP/Bohlin2013/GO12813.html}}. These flux standards are claimed to be at the level of $\sim 1$\% photometric accuracy. In addition a B-type telluric standard HIP 38858 was observed to separately correct for the OH and H$_2$O telluric absorption features in the red. The separation of these features by molecular species allows for a more accurate telluric correction which accounts for night to night variations in the column density of these two species.

All the data cubes were reduced using the  \textsc{pywifes} \footnote {http://pywifes.github.io/pipeline/} data reduction pipeline (Childress et al. 2014) using the procedures described therein. In particular, the photometric calibration of the WiFeS instrument is very reliable, since the instrument is based upon image slicing rather than optical fibre technology, and the spatial format is fixed. The spaxel to spaxel sensitivity variations are calibrated using both internal and twilight flat fields, and the spectral sensitivity from the multiple standard star observations. From the sensitivity function derived from the three independent standards observed over the two nights we estimate the photometric calibration of the instrument to be better than 3\% in the wavelength range 3700 to 6800 {\AA}.

\begin{figure*}[htb!]
\includegraphics[scale=0.6]{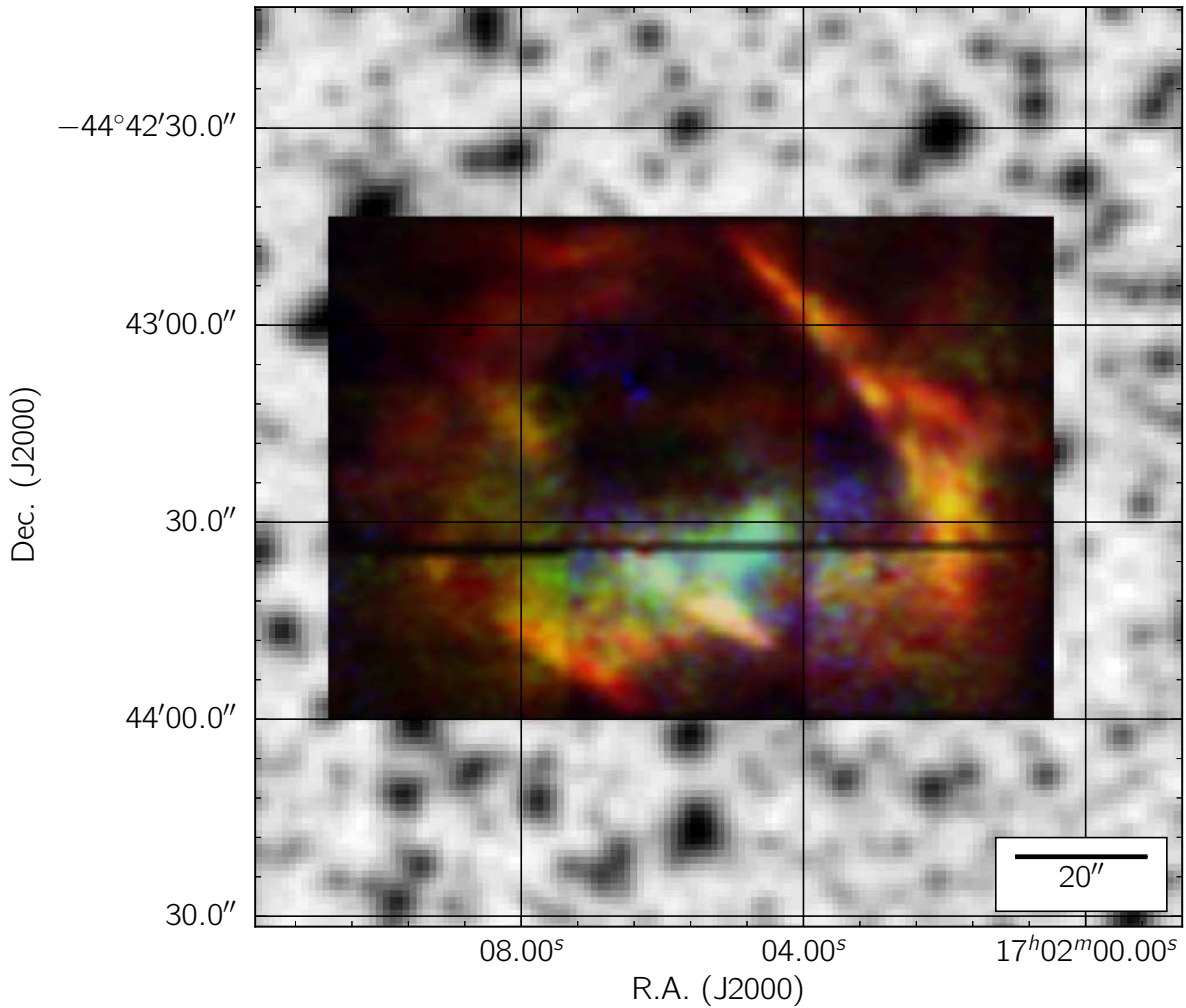}\label{fig:image}
\caption{The  WiFeS emission line map of the PN G342.0-01.7 superimposed on the  DSS-2 (red) stellar field. The image is a mosaic of the  9 observed positions on the nebula, where each frame has a field of view 25\arcsec $\times$ 38\arcsec. The colour image is a composite of [NII] (red) H$\alpha$ (green) and [OIII] (blue), and reflects the excitation structure of the nebula. The horizontal gap results from a pointing offset error between the first and second night's observations.}
\end{figure*}

PN G342.0-01.7 has an angular size of $90.0^{\prime\prime}\times63.0^{\prime\prime}$ square arcsec. Observing the full spatial extent of this object required a mosaic of $3\times3=9$ WiFeS fields with 1-2 arcsec spatial overlap between each field. Each individual field was observed $2\times600s=1200s$ on-source, simultaneously with the red and the blue arm of WiFeS.

Each frame was reduced individually using \textsc{pywifes} into a flux and wavelength calibrated 3-D data cube. The final mosaic was assembled using a custom \textsc{python} script (available on demand). Because the World Coordinate System (WCS) information in the data \textsf{FITS} headers can only be trusted to a level of a few arcsec, the H$\alpha$ and H$\beta$ integrated flux maps for each frame are constructed and used to identify the best spatial shifts between each two fields. Given the seeing conditions during the observations (1.5$^{\prime\prime}$-2$^{\prime\prime}$), we round the spatial shifts to the nearest integers for simplicity, and to avoid superfluous resampling of the data. All data cubes were forced on the same wavelength grid in the data reduction process, so that no shift is required along that axis.

The final alignment of the different fields were checked against the Digitized Sky Survey (DSS) 2 red band image of the area \footnote{The DSS-2 image was obtained from the ESO Online Digitized Sky Survey: {\tt http://archive.eso.org/dss/dss}}. The final accuracy of the spatial alignment of each individual field is of the order of 1 arcsec. In particular, the DSS-2 image was essential for combining the fields acquired during the first and the second night of observations, which are separated by a gap of 1-2 arcsec as a result of a pointing offset between the two night's observations.

 A composite [NII] (red) H$\alpha$ (green) and [OIII] (blue) colour image of the mosaic is shown in Figure 2, where North is up and East to the left. This image goes much deeper and is of higher (spatial) resolution compared to that given in the MASH catalogue.  The choice of the colors is to emphasise the excitation structure of the nebula since [O III] is confined to the H$^+$ and He$^+$ zone, while [N II] emission is found only in the  H$^+$ and He$^0$ zone of the nebula.

 It is apparent that the peak of excitation is displaced from the geometrical centre of the nebula, at about RA = 17h 02m 05s; Dec = 44$^o$ 43' 31'' (J2000). In addition, the nebula surface brightness is enhanced around this excitation peak. Overall, the displacement of the excitation centre towards the SW (determined from the [O\,{\sc\small III}] image), the enhancement of the surface brightness in the SW quadrant, the overall morphology including the faint halo extending to the NE seen in the ESO image (Figure \ref{fig:FORS2}) are all suggestive of an interaction with the surrounding interstellar medium (ISM), with the nebula moving in the SW direction relative to its ISM.

 \begin{figure}[htb!]
\includegraphics[scale=0.32]{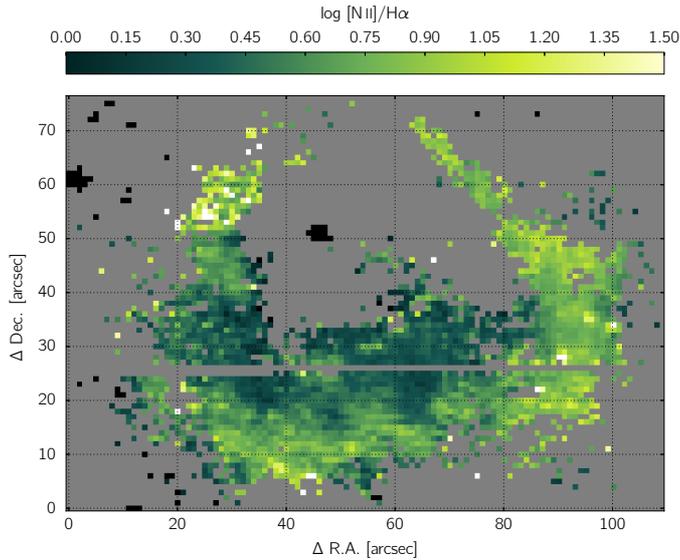}\label{fig:NII/Ha}
\caption{The spaxel to spaxel variation of the [N\,{\sc\small II}] /H$\alpha$ ratio within PN G342.0-01.7. A signal to noise cut of 10 has been applied. The strongest  [N\,{\sc\small II}]  emission relative to H$\alpha$ is clearly located in the brightest outer filaments, and in the outer parts of the nebular shell.}
\end{figure}

 The details of the excitation structure are further revealed in a map of the [N\,{\sc\small II}] /H$\alpha$ ratio, shown in Figure 3. We see strong [N\,{\sc\small II}]  emission relative to H$\alpha$  located in the brightest outer filaments, and in the outer parts of the nebular shell, while the lowest  values of the  [N\,{\sc\small II}] /H$\alpha$ ratio occur towards the centre. The signal to noise in the blue part of the spectrum is too low to enable us to provide similar maps of blue to red line ratios such as H$\alpha$/ H$\beta$, so we are unable to trace the detailed spaxel to spaxel variations in interstellar reddening.

\section{Derived Global Nebular Parameters} \label{Props}
\subsection{Nebular spectrum and reddening}
The global spectrum of PN G342.0-01.7 was extracted from the mosaiced data cubes of the blue and the red spectra. The scaling between the blue and red spectra was improved by measuring the total flux in the common spectral range between the blue and red channels (5500-5700 {\AA}), and applying a 2\% scale factor to the red spectrum to make the two measured fluxes match. The integrated nebular fluxes were derived from the mosaiced image after subtracting the stellar background stars.  The resultant global spectrum is shown in Figure 4. Note that high interstellar dust reddening, and the residual blue stellar continuum from faint stars in the field makes it more difficult to accurately measure fluxes below $\lambda \sim 4200$\AA .

The line fluxes and their associated uncertainties were measured in IRAF {\tt splot} \footnote{IRAF is distributed by the National Optical Astronomy Observatory, which is operated by the Association of Universities for Research in Astronomy (AURA) under a cooperative agreement with the National Science Foundation.} task. Line fluxes were integrated between two given limits, over a local continuum fitted by eye, using multiple Gaussian fitting for the line profile.

In general, the uncertainty of the measured emission lines fluxes in the red is smaller than the blue since the blue lines are much fainter due to the heavy reddening of this object. The first four emission lines in Table 1, have measurement uncertainties between 10\%-20\%. For other lines, the uncertainties is less than 6\% for lines stronger than $H\beta$ and less than 12\% for lines weaker than $H\beta$. The uncertainties in the form of the reddening function increase the measurement errors considerably, particularly for the blue lines. We have estimated these errors by using various forms of reddening functions from the literature with total to selective extinctions in the range $3.0 < R < 4.6$.

We have used the Nebular Empirical Abundance Tool (NEAT \footnote{The code, documentation, and atomic data are freely available at http://www.sc.eso.org/~rwesson/codes/neat/.}; Wesson et al. 2012) for plasma diagnostics. From the emission lines list, NEAT implemented a standard techniques to estimate the interstellar reddening, electron temperature and density, ionic and elemental abundances. The reddening coefficient $c(H\beta)$ was determined from the ratios of hydrogen Balmer lines, in an iterative procedure. A preliminary reddening value was derived assuming $H\alpha$, $H\beta$, $H\gamma$ line ratios at electron temperature ($T_e = 10000$ K) and electron density ($N_e = 100$ cm$^{-3}$). NEAT uses a Monte Carlo technique to propagate uncertainties from line flux measurements to elemental abundances.

Here, we select the (default) Galactic extinction law of Howarth (1983), which is a functional characterization of the one from Seaton (1979). We obtained a reddening coefficient $c(H\beta)= 1.81$, E(B-V)= 1.24 and a total $H\beta$ flux $\log F(H\beta) = -12.88$. The line ratios [O\,{\sc\small III}] $\lambda 5007/\lambda 4959$ and [N\,{\sc\small II}] $\lambda 6584/\lambda 6548$ both measured to be close to 3.0, in good agreement with theoretical predictions of 2.86 and 3.0 respectively (Storey \& Zeippen 2000). Table 1 lists the observed and dereddened line strengths of the object. The first two columns give the emission lines identification with their laboratory wavelengths, while columns 3 and 4 give the observed F($\lambda$) and the dereddened line strengths I($\lambda$), scaled to $H\beta =100$.

\begin{figure}
\includegraphics[width=9cm,height=9cm]{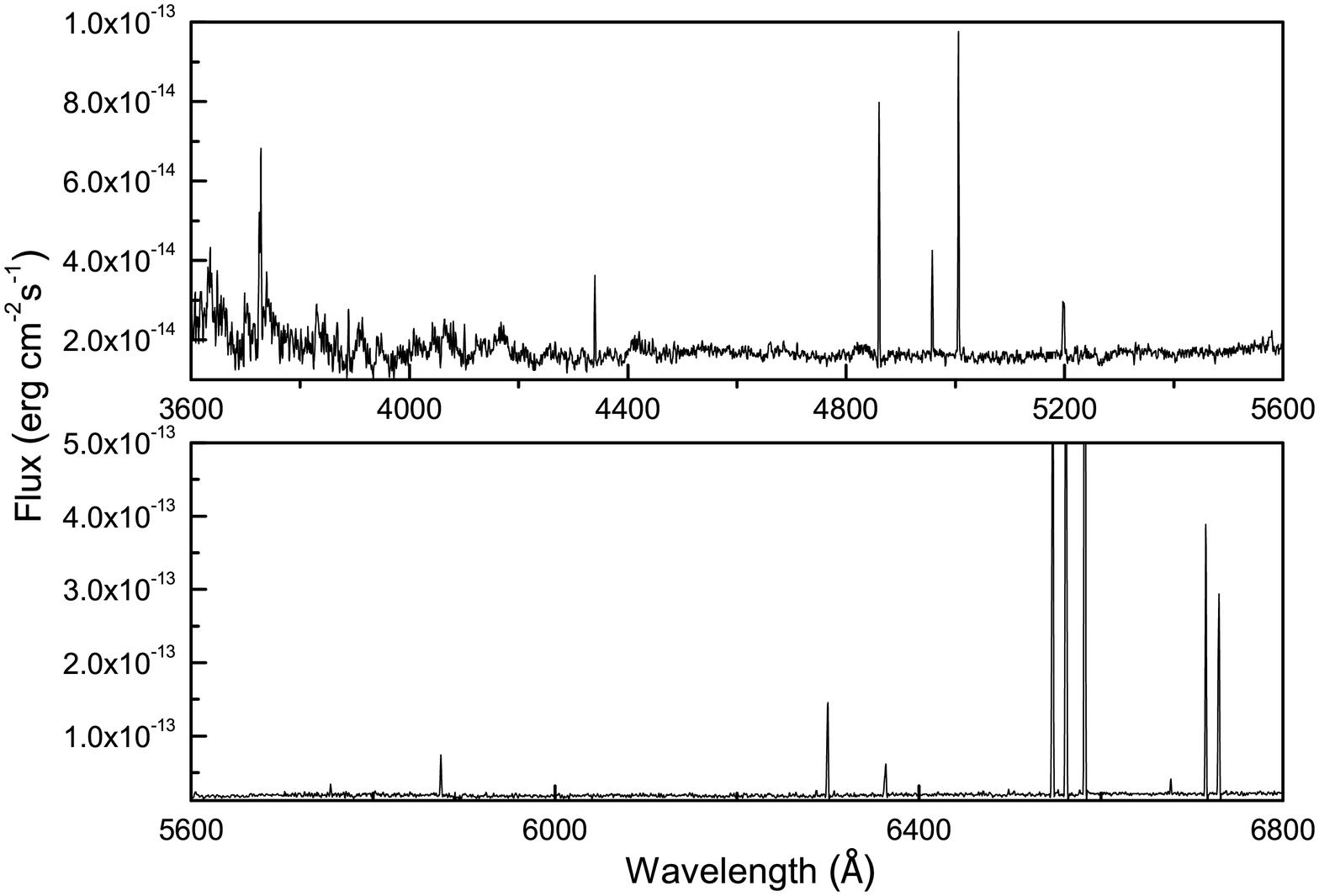}
\vspace{0.1cm}
\caption{The blue and red optical integrated spectrum of PN G342.0-01.7. Note that residual stellar continuum in the field limits the S/N of the line detection below $\lambda \sim 4300$. }
\end{figure}

\begin{table}\label{Table1}
\centering
 \small
   \caption{Observed $F(\lambda)$ and dereddend $I(\lambda)$ line intensities on a scale $F(\rm H\beta) = 100$. The errors on the de-reddened fluxes include the uncertainty associated with the assumed form of the interstellar reddening function.}
  \begin{tabular}{llll}
 \hline \hline
  $ \lambda_{lab} $ & Ion & $F \left( \lambda \right) $ & $I \left( \lambda \right) $ \\
  \hline
 3726.03 & [O~{\sc ii}]    &  34$\pm$  6 &  100$\pm$  25 \\
 3728.82 & [O~{\sc ii}]    &  59$\pm$  8 &  172$\pm$  35 \\
 4101.74 & H$\delta$       &  14.2$\pm$  2.7 &   29$\pm$ 7 \\
 4340.47 & H$\gamma$       &  32.2$\pm$  2.0 &   54.6$\pm$  5.0 \\
 4861.33 & H$\beta$       &  100.0$\pm$  1.5 &  100.0 $\pm$  1.5 \\
 4958.91 & [O~{\sc iii}]   &  41.0$\pm$  1.0 &   37.2$\pm$   0.8 \\
 5006.84 & [O~{\sc iii}]   &  129.6$\pm$  1.5 &  111.7$\pm$   2.8 \\
 5200.26 & [N~{\sc i}]     &  33.4$\pm$  1.4 &   23.7$\pm$   1.1 \\
 5754.60 & [N~{\sc ii}]    &  16.8$\pm$  1.8 &    7.5$\pm$   0.9 \\
 5875.66 & He~{\sc i}      &  62.5$\pm$  2.1 &   25.6$\pm$   1.5 \\
 6300.30 & [O~{\sc i}]     &  177.5$\pm$  1.8 &   54.9$\pm$   5.0 \\
 6363.77 & [O~{\sc i}]     &  62.2$\pm$  1.2 &   18.5$\pm$   1.7 \\
 6548.10 & [N~{\sc ii}]    &  1108.9$\pm$  2.2 &  295 $\pm$   25 \\
 6562.77 & H$\alpha$       &  1120.0$\pm$  1.8 &  295$\pm$   25 \\
 6583.50 & [N~{\sc ii}]    &  3427.4$\pm$  2.1 &  893$\pm$   90.0 \\
 6678.16 & He~{\sc i}      &  20.6$\pm$  1.0 &    5.1$\pm$   0.6 \\
 6716.44 & [S~{\sc ii}]    &  419.4$\pm$  1.0 &  101.2$\pm$   12.5 \\
 6730.82 & [S~{\sc ii}]    &  298.4$\pm$  1.1 &   71.4$\pm$   9.0 \\
 \hline
 \end{tabular}
\end{table}

\subsection{Excitation class}
Three methods are known in the literature to determine the excitation class (EC) of planetary nebula, which generally is a good indicator of the temperature of the central star (Gurzadyan 1988; Dopita et al. 1992; Reid \& Parker 2010). At low excitation class (up to about 5) the [O\,{\sc\small III}] $\lambda 5007/H\beta$ line ratio is the key indicator, while above this the strength of the He II emission at  4686 {\AA} provides the best indication
of excitation class of the nebula. This line is only present when the central star has $T_{\rm eff} > 85000$K. In G342.0-01.7 the low  [O\,{\sc\small III}] $\lambda 5007/H\beta$ line ratio, the high [N\,{\sc\small II}] /$H\alpha$ line ratio, and the absence of the HeII 4686 line would indicate that the object belongs to the very low excitation class EC = 0.5, and is also likely to show evidence of CN-processed dredged-up material leading to an enhancement of N and He in the nebula. The effective temperature indicated by the excitation class is only $\rm T_{\rm eff} \sim 35000 K$ (Dopita et al. 1992). However, as we will show below, for evolved PNe such as the one of the current study, this naive classification by excitation class can be quite misleading.

\subsection{Ionic and elemental abundances}
From the spectral data of Table 1, we can in principle derive the electron density from the  [S~II] $\lambda\lambda$6716/6731 line ratio and estimate the temperature  from the [N~II] ($\lambda 6548 + \lambda 6584$)/$\lambda$5754 line ratio. However, the [S~II] line ratio is found to be close to its low density limit $N_e ([S~II]) = 16$ cm$^{-3}$, and so does not provide a meaningful constraint on the density. The [N II] lines provide an estimate of the mean electron temperature of $T_e ([N~II]) = 8200$ K for the whole PN.

A first estimate of the ionic and total element abundances of PN G342.0-01.7 was determined by the NEAT. Optical recombination lines (ORL) were used to derive ionic abundances for helium, while Collisional excitation lines (CEL) were used to derive ionic abundances of nitrogen, oxygen and sulfur. The total elemental abundances were derived applying the ionization correct factors (ICF) of Kingsburgh \& Barlow (1994), to correct for unseen ions. The reddening coefficient, nebular temperature and density, ionic and elemental abundances and their associated errors are presented in Table 2.  The results indicate a pronounced enrichment in both the helium and nitrogen abundances (He/H = 0.16 and N/O = 1.0). Peimbert (1978) has defined the Type I PNe as the objects with He/H $> 0.14$ or N/O $> 1$. Later on Peimbert \& Torres-Peimbert (1983) relaxed the definition by considering those objects with He/H $> 0.125$ or N/O $> 0.5$. More rigid criteria to define Type I PNe were given by Maciel \& Quireza (1999), which led to the definition that those objects satisfy the conditions He/H $> 0.125$ and N/O $ > 0.5$ should be classified as Type I PNe.

The excess in He and N nebular abundances of Type I PNe refer to the objects that underwent hot-bottom burning and dredge-up of CN-processed material into the envelope of the precursor star. This process is restricted to the more massive progenitor stars in the intermediate mass range, which are therefore associated with the youngest Galactic stars found in the galactic thin disk ($z \leq 300$ pc, Gilmore \& Reid 1983).

The Type I classification of PN G342.0-01.7 on the basis of the  chemical abundances derived in Table 2 was confirmed in light of its kinematic properties. Applying the derived PN distance (see section 3.5) the object lies at a vertical height $z \simeq 61$ pc below the Galactic Plane. In addition it has a small peculiar velocity {\bf ($|\Delta V|$)} defined as the difference between the observed local standard of rest radial velocity and the velocity determined from the rotation curve. We find $\sim |\Delta V| = 6.9$ km $s^{-1}$. Maciel \& Dutra (1992) show that objects belonging to the Galactic thin disk should have  $|\Delta V| \leq 60$ km $s^{-1}$. Therefore we may conclude that PN G342.0-01.7 belongs to the Galactic thin disk consistent with its Type I classification.

\begin{table}\label{Table2}
  \caption{Ionic and elemental abundances of PN G342.0-01.7.}
  \begin{tabular}{lc}
\hline
 \vspace{0.1cm}\\\multicolumn{2}{l}{Extinction}\\ \hline
c(H$\beta)$:                        & ${  1.81}\pm{  0.02}$ \\
\vspace{0.1cm}\\\multicolumn{2}{l}{Nebular density and temperature}\\ \hline
{}[S~{\sc ii}] density             & ${ 16.24}\pm{  10}$ \\
{}[N~{\sc ii}] temperature          & ${ 8190}\pm{  300}$ \\
 \vspace{0.1cm}\\\multicolumn{2}{l}{ORL abundances}\\ \hline
He$^{+}$/H                          & ${  0.16}\pm{  4.78\times 10^{ -3}}$ \\
 He$^{2+}$/H                        & -- \\
He/H & ${  0.16}\pm{  4.78\times 10^{ -3}}$ \\
 \vspace{0.1cm}\\\multicolumn{2}{l}{CEL abundances}\\ \hline
N$^{+}$/H                           & ${  2.55\times 10^{ -4}}^{+  2.79\times 10^{ -5}}_{ -3.62\times 10^{ -5}}$ \\
 N$^{2+}$/H                         & -- \\
 N$^{3+}$/H                         & -- \\
 N$^{4+}$/H                         & -- \\
icf(N)                              & ${  1.31}^{+  0.03}_{ -0.03}$ \\
N$^{}$/H                            & ${  3.51\times 10^{ -4}}^{+  4.02\times 10^{ -5}}_{ -3.61\times 10^{ -5}}$ \\
O$^{0}$/H                           & ${  2.46\times 10^{ -4}}^{+  4.23\times 10^{ -5}}_{ -3.61\times 10^{ -5}}$ \\
O$^{+}$/H                           & ${  2.65\times 10^{ -4}}^{+  6.21\times 10^{ -5}}_{ -5.03\times 10^{ -5}}$ \\
O$^{2+}$/H                          & ${  7.89\times 10^{ -5}}^{+  1.14\times 10^{ -5}}_{ -1.36\times 10^{ -5}}$ \\
 O$^{3+}$/H                         & -- \\
icf(O)                              & ${  1.00}\pm{  0.00}$ \\
O$^{}$/H                            & ${  3.49\times 10^{ -4}}^{+  7.40\times 10^{ -5}}_{ -6.11\times 10^{ -5}}$ \\
S$^{+}$/H                           & ${  6.37\times 10^{ -6}}^{+  6.85\times 10^{ -7}}_{ -7.62\times 10^{ -7}}$ \\
 S$^{2+}$/H                         & -- \\
icf(S)                              & ${  3.67}^{+  0.25}_{ -0.26}$ \\
S$^{}$/H                            & ${  2.45\times 10^{ -5}}^{+  2.79\times 10^{ -6}}_{ -2.51\times 10^{ -6}}$ \\
  \hline
\end{tabular}

\end{table}

\subsection {The search for a central star candidate}
We inspected the digitized UKST (Blue \& Red) plates from the SuperCOSMOS sky survey, to search for a blue central star (CS) candidate. Only one possible CS candidate was found which is brighter in the blue plate relative to the red plate when compared to the field stars. This lies at $\alpha =$ 17:02:4.388, $\delta =$ -44:43:24.74 (J2000), and  is located very close to the geometric center of the nebula ($\alpha =$ 17:02:4.3, $\delta =$  -44:43:20). Although it lies close to a much brighter star, we were able to extract a fairly clean spectrum of this object from the WiFeS data cube. This is shown in Figure 5. From this spectra we conclude that the star is a foreground field A-Type dwarf, and is therefore unlikely to be associated with the PN.
\begin{figure}
\includegraphics[height=4cm]{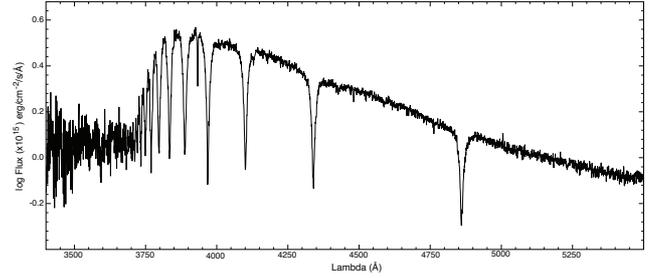}
\caption{The spectrum of the central star candidate of PN G342.0-01.7 extracted from the WiFeS data cube. We conclude that this is a field A-Type dwarf star, not related to the PNe.}
\end{figure}

As we will show below, the central star has a very high temperature, is very reddened, and therefore would remain undetectable in these observations.

\subsection {Distance and $H\beta$ luminosity}
The determination of the distance of a planetary nebula is the fundamental parameter which permits an estimation of most nebular characteristics. However,  finding  a reliable distance for the PN is a long-standing problem. Generally, two major techniques
are used, either individual estimates based upon, for example, the determination of both $\log T_{\rm eff}$ and $\log g$ of the central star. Failing such direct techniques, we must instead rely on statistical methods. A summary on the applications, assumptions, limitations, and uncertainties  of these methods was given in Ali et al. (2015) \& Frew et al. (2015).

Many of the statistical methods are based on an empirical relationship between two distance-dependent nebular parameters (e.g. Van de Steene \& Zijlstra 1995; Stanghellini et al 2008; Ali et al. 2015). All these methods are rely on the observed 5GHz radio flux. Since there is no radio measurement for PN G342.0-01 at 5GHz, we have used Equation (6) in Cahn et al. (1992) to convert the total $H\beta$ flux into 5GHz radio flux. Using the derived 5GHz radio flux (30.06 mJy) and the nebular angular radius (38.25 arcsec,  Parker et al. 2006) and following the statistical distance scale of Ali et al. (2015), we derived a nebular distance of $2100 \pm 630$ pc.

Frew et al. (2015) introduced a linear empirical relationship between the nebular H$\alpha$ surface brightness and nebular physical radii ($S_{H\alpha}-r$) with a slope between -3.3 and -3.8 depending on the calibrated subset used.  They found that optically thick PNe tend to populate the upper boundary of the trend with a mean dispersion of $\pm 28\%$, while optically thin PNe fall along the lower boundary with a mean dispersion of $\pm 18\%$ in the $S_{H\alpha}-r$ plane. Applying the definition of optically-thick PNe have relatively strong low excitation features such as [N II], [O II] and [S II] lines (Kaler \& Jacoby 1989 and Jacoby \& Kaler 1989), and hence the flux ratio of [N II] $\lambda 6584/H\alpha \geq 1$, we were followed the optically thick form of $S_{H\alpha}-r$ relation (Equation 17; Frew et al. 2015).
From the spectrum analysis, we derived a total $H\alpha$ flux log F$(H\alpha) = -11.77$ erg cm$^{-2}$ s$^{-1}$. This value was converted into a reddening corrected $H\alpha$ surface brightness log S$(H\alpha) = -3.55$ erg cm$^{-2}$ s$^{-1}$ sr$^{-1}$, applying Frew et al. (2015) procedure. This surface brightness yields a distance of $2022 \pm 566$ pc.
We have to take the uncertainty in the calculated distance with caution, because it is derived from the dispersion in the calibration samples used in both distance scales. Consequently the uncertainty in the distance could be significantly larger.

Finally we adopt a nebular average distance of $2060\pm 600$ pc. On the basis of this distance, we were derived an absolute luminosity of $L_{\rm H\beta}= 4.2\pm 2.5\times10^{33}$~erg~s$^{-1}$, and a physical dimension of $0.90\times0.63$~pc.

\subsection {Radial and expansion velocities}

We have used the red channel of high resolution spectra to estimate the radial velocity. Five emission lines of $H\alpha$, [N~II], and [S~II] in the red spectrum were used to find the observed radial velocity $RV_{\rm obs}$ of the object. We determined $RV_{\rm obs} = -57\pm1.7$ km s$^{-1}$. This leads to a derived heliocentric radial velocity $RV_{\rm hel} = -32.7\pm1.7$ km s$^{-1}$. The correction to local standard of rest was performed using the appropriate $RV_{\rm hel}$ of the solar motion. We estimate $RV_{\rm LSR} = -27.7\pm1.7$ km s$^{-1}$ for this object.

The expansion velocity of a PN is a key parameter in determining its evolution. The well defined emission lines [N~II], [S~II] and H$\alpha$ were used to determine the expansion velocity of the PN G342.0-01.7. Following, for example, Gieseking et al. (1986):
\begin{equation}\label{3}
  V_{\rm exp} = 0.5 \left[w^2_{\rm obs} - w^2_{\rm inst} - 8(\ln 2)kT_e/m\right]^{1/2}
\end{equation}
 where $w_{\rm obs}$ is the observed FWHM of the measured line, $w_{\rm inst}$ is  the instrumental FWHM (see Section 2.2), $k$ Boltzmann's constant, $T_e = 8190$K is the electron temperature, and $m$ the atomic mass of measured line. The observed FWHM of each line was determined using the IRAF {\tt splot} task. We determined $V_{\rm exp} {\rm [N~II]} = 18.8$ km~s$^{-1}$,  $V_{\rm exp} {\rm [S~II]} = 21.2$ km~s$^{-1}$, and $V_{\rm exp} [{\rm H\alpha}] = 20.6$ km~s$^{-1}$. Therefore, we adopt a mean expansion velocity of $V_{\rm exp} = 20.2\pm 1.3$ km~s$^{-1}$ for this PN.

\begin{table}
  \caption{The derived global parameters for PN G342.0-01.7.}
  \begin{tabular}{ll}
     \hline
     \hline
  Distance & 2.06 kpc \\
  Angular size & 90.0\arcsec x 63.0\arcsec \\
               & Parker et al. (2006) \\
  Physical size & 0.90 x 0.63 pc \\
  Volume density &$\lesssim 20$ cm$^{-3}$ \\
  Dynamical age & $\gtrsim $ 20000 yr \\
  $RV_{LSR}$ & -27.7$\pm$1.7 km s$^{-1}$ \\
  $V_{exp}$  & 20.2$\pm$1.3 km $s^{-1}$\\
  $|\Delta V|$ &  $\sim 7$ km $s^{-1}$\\
  Excitation class & 0.5 \\
  Peimbert class & Type I \\
  E(B-V) &   1.24 \\

  \hline
\hline
\end{tabular}
\end{table}

\subsection{Nebular age}
Provided that we can treat the nebula as being in free expansion at $V_{exp} =20.2$\,km s$^{-1}$,  the observed size leads to a dynamical age of the nebula of $\sim 20000$\,yr. Given the evidence presented in Section \ref{Interaction}, the nebula is most likely no longer in free expansion, and therefore these ages of the nebular shell  should be treated as a lower limit. The ionized mass, density and dynamical age of the object all support the idea that we are dealing with a considerably evolved planetary nebula. A summary of the derived nebular properties is given in Table 3.

\subsection{Evidence for interaction with the ISM}\label{Interaction}
The  PN G342.0-01.7 shows three distinct characteristics of an on-going interaction with the surrounding ISM. First, the morphology of  PN G342.0-01.7 as seen in Figure 1 is that of an elliptical ring with a brightness enhancement on the southwest side. In addition three straight filaments (or stripes) cross the PN in the NE-SW direction, and all are brightest in the SW part of the nebula. Such features have been seen before in other PNe, e.g. PN G218.9-10.7 (Ali \& Pfleiderer 1999) and NGC 6894 (Soker \& Zucker 1997), where they are also attributed to the interaction between the PN and its surrounding ISM. In particular, stripes arise due to the stripping of the object's halo by ISM (many examples of such objects were discussed by Ali et al. 2012) although Soker \& Zucker (1997) speculate that such stripes are shaped by the interstellar magnetic field due to their orientation.

Also evident is a faint outer halo, extended to some considerable distance in the NE direction. These features are consistent with ram-pressure stripping of the PNe shell on its trailing edge.

Second, the excitation centre of the PNe as traced by the  [O\,{\sc\small III}] $\lambda 500$7 emission in Figure 2 clearly shows a peak near RA = 17h 02m 05s; Dec = 44$^o$ 43' 31'' (J2000). This peak (assumed to be associated with the nearby presence of the exciting star) is strongly displaced from the geometrical centre of the main ring of the PNe, which is located near RA = 17h 02m 06s; Dec = 44$^o$ 43' 45'' (J2000). This suggests that the expansion of the nebula has been impeded in the SW direction, towards the assumed direction of motion of the PNe through its surrounding ISM. The brightening of the filaments in the S - SW quadrant of the nebula would therefore be ascribed to both an intrinsically higher pressure or nebular density, as well as a reduced distance from the exciting star, with consequently a stronger radiation field.

Third, the filaments in the SW quadrant are characterised by an anomalously high  [O\,{\sc\small I}] $\lambda 6300$/H$\alpha$ ratio. As discussed below, we associate this with the presence of a slow reverse shock propagating into the PNe ejecta caused by the ram-pressure of the motion of the PNe through the surrounding ISM.

\section{Theoretical Modelling} \label{Model}
\subsection{Photoionisation Model}
We have used the {\tt MAPPINGS~V} code to construct a detailed photoionisation model of PN G342.0-01.7. The {\tt MAPPINGS~V} code is a major upgrade of the {\tt MAPPINGS~IV} code described in Dopita et al. (2013), and the new physics included in this new version will be described in Sutherland et al. (2015, in preparation). This code provides single framework within which to model the emission line and continuum spectra of equilibrium ionisation objects such as HII regions or PNe while at the same time being able to do the same for plasmas which are well out of collisional or photoionisation equilibrium such as the radiative shocks in supernova remnants (SNR) or in the Herbig-Haro objects. In the models described here we have used the non-LTE model atmospheres described by Rauch (2003) taken from the solar abundance ratio grid, and with $\log g = 7$ and 8. Where the requested $T_{\rm eff}$ is not computed, {\tt MAPPINGS~V} makes a $T^4$ interpolation between available models at each frequency to obtain the ionising spectrum of the central star to be used in the modelling.

Our model aims to match the global spectrum, the size, and the absolute H$\beta$ luminosity of the PNe. Since we do not know, \emph{a priori}, the effective temperature of the central star, we ran many models of different effective temperature. Surprisingly, we could not get a satisfactory model for any stars with $\log T_{\rm eff} < 4.8$, although the excitation class of EC=0.5 would be expected to yield $\log T_{\rm eff} \sim 4.5$. This difference seems to arise as a result of the evolved nature of PNe, and the relatively low density \emph{and} ionisation parameter characterising the photoionised plasma. The ionisation parameter can be constrained from the observed [O~III]/[O~II]  ratio, although for these models the variation with ionisation parameter is relatively weak. We find a photon field averaged ionisation parameter $\log U = -3.4$ in our best-fit model. The central star temperature is constrained in the range $4.95 < \log T_{\rm eff} < 5.08$. The principle diagnostics of the effective temperature are the [O\,{\sc\small III}] $\lambda 5007/H\beta$ line ratio which increases with $T_{\rm eff}$, and the absence of detectable He\,{\sc\small II} $\lambda 4686$ emission, which appears at detectable levels in the models when $\log T_{\rm eff} > 5.1$.  In our best-fit model presented below we adopt $\log T_{\rm eff} = 5.05$, although other models in the temperature range defined above are almost equally acceptable.

The luminosity of the central star is constrained by the requirement that the model reproduce the observed H$\beta$ flux ($L_{\rm H\beta}= 4.2 \times10^{33}$~erg~s$^{-1}$ for an assumed distance of 2.06 kpc.  The pressure in the (isobaric) models is adjusted until the correct overall dimension of the PNe is achieved. This gave $\log P/k =5.85$\,cm$^{-3}$K. For the best fit model, $L/L_{\odot} = 118$ for or an assumed distance of 2.06 kpc. These luminosities, the high $T_{eff}$ of the central star, and the very heavy reddening ensure that the central star will remain undetectable at visual wavelength. We estimate it to have a visual magnitude of $\sim 22.0$.

The inferred position of the central star is shown in Figure 4 relative to the H-burning evolutionary track of Vassiliadis \& Wood (1993). On this figure we have also plotted isochrones for the evolutionary age dated from the time where the central star achieved a temperature of $\log T_{\rm eff} = 4.0$. It is clear that the inferred position of the central star of PN G342.0-01.7 is reasonably consistent with the inferred dynamical age of $t  \geq 20000$~yr. However, the uncertainty in the derived stellar parameters precludes any estimate of the initial mass of the central star. Nonetheless, we can confirm both the evolved nature of the object, and its advanced age.

\begin{figure}
\includegraphics[scale=0.6]{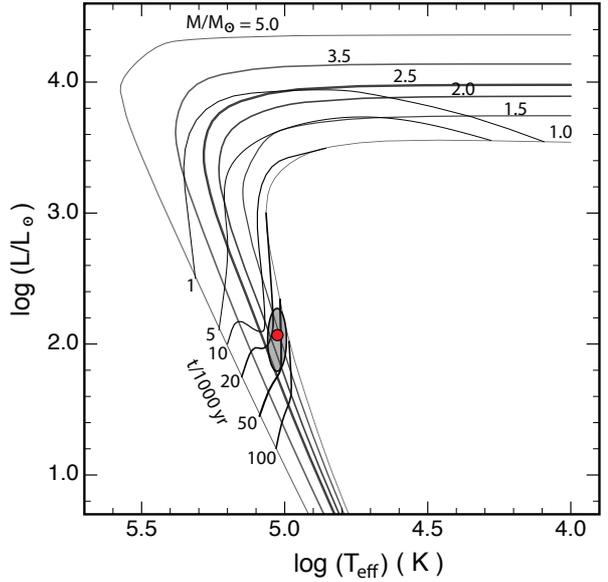}\label{HRDiag}
\caption{{\bf The position of the exciting star of PN G342.0-01.7 is marked as a red circle as derived from photoionisation modelling compared with evolutionary tracks of central stars of PNe from Vassiliadis \& Wood (1993).  The gray area represents the error ellipse defined by the uncertainties in the flux and distance estimates, and by the uncertainties in the theoretical modelling. Note that the tracks are marked in terms of the initial stellar mass, rather than of the remnant core.}}
\end{figure}

As described above, the pressure in the (isobaric) model is determined by the requirement that it reproduce the observed size of the nebula. This is achieved for $\log P/k = 5.85$~cm$^{-3}$K, for the assumed distance of 2.06 kpc. From basic Str\"omgren theory, the nebular pressure required to match the observed diameter scales with assumed distance as $D^{-1/2}$. This pressure yields a mean electron density of $n_e = 26$~cm$^{-3}$ in the ionised plasma. The mean electron temperature in the [N~II]-emitting zone in the model is predicted to be 10000K, appreciably hotter than that observed, 8200K, although this latter figure is subject to quite large errors, given the faintness of the [N II] $\lambda 5755$ line.

The line flux predictions of our best-fit photoionisation model are summarised in Table 4 (column 5). This confirms the Type I nature of this PNe, with a large over-abundance of both He and N being derived. In dusty PNe models we must be careful to distinguish between the gas-phase and total abundances of the elements.  The gas-phase abundances are what are fitted by a code such as NEAT, while the photoionisation model uses the total abundances of the heavy elements, accounting for the fraction of these that are locked up in dust according to a standard dust model. The depletion factors of the heavy elements onto dust which we use are derived from the formulae of {Jenkins, 2009}, extended to the other elements on the basis of condensation temperatures and/or position on the periodic table. To scale the abundance set to include dust we must adopt a base Fe depletion in the nebular gas,  here taken to be $\log D_{\rm Fe} \sim -1.5$.

In our model, we have He/H = 0.155, N/H $= 2.3\times10^{-4}$ (total), N/H $= 2.1\times10^{-4}$ (gas phase), N/O $= 1.42$  (total), N/O $= 1.67$  (gas phase), O/H $= 1.6\times10^{-4}$  (total),  O/H $= 1.3\times10^{-4}$  (gas phase), and S/H $= 7.5\times10^{-6}$  (total and gas phase). These figures should be compared with those derived directly from the one-zone spectrum analysis in Table 2. Generally speaking, the agreement is satisfactory, with the notable exception of O, for which the model gives a much lower abundance than the one-zone analysis. The cause of this can be found in the [O~I] zone. The photoionisation model predicts much weaker emission than the observations provide. As we will show in the next section, this is a symptom of the presence of reverse shock  emission.
\begin{table*} \label{Table4}
\vspace{1cm}
\centering
  \caption{Observed line Intensities for PN G342.0-01 compared with our photoionisation and shock models.}
  \begin{tabular}{llcccc}
     \hline
     \hline
         &   & (3) & (4) & (5) & (6) \\
$\lambda$	 & Line & Observed	& PNe + Shock	& PNe & Shock  \\
  	                   & ID &  (H$\beta$=100)	& (13\% shock)	& Model & Model \\
\hline
3728 & [O~II] & $273\pm 25$ & 296 & 341 & 7.9\\
4101 & H$\delta$ & $29.0\pm7.0$ & 23.8 & 25.7 & 9.9\\
4340 & H$\gamma$ & $54.6\pm 5.0$ & 45.7 & 46.7 & 39.0\\
4861 & H$\beta$ & $100\pm 1.5$ & 100.0 & 100 & 100\\
4959 & [O~III] & $37.2\pm 0.8$ & 40.6 & 47 & 0.0\\
5007 & [O ~III] & $112\pm 2.8$ & 117 & 136 & 0.0\\
5199 & [N~I] & $23.7\pm 1.1$ & 23.1 & 6.9 & 127\\
5755 & [N~II] & $7.5\pm 0.9$ & 14.4 & 16.5 & 0.4\\
5876 & He~I & $25.6\pm 1.5$ & 22.4 & 25.9 & 0.2\\
6300 & [O~I] & $54.9\pm 5.0$ & 54.8 & 24.6 & 248\\
6363 & [O~I] & $18.5\pm 1.7$ & 17.5 & 7.9 & 79.5\\
6548 & [N~II] & $295\pm 25$ & 290.5 & 291.5 & 15.2\\
6563 & H$\alpha$ & $295\pm 25$ & 318.5 & 291.5 & 490\\
6584 & [N~II] & $893\pm 90$ & 854.8 & 981 & 44.8\\
6678 & He~I & $5.1\pm 0.6$ & 6.5 & 7.3 & 1.1\\
6717 & [S~II] & $101.2\pm 12.5$ & 116 & 102.7 & 201.5\\
6731 & [S~II] & $71.4 \pm 9.0$ & 81.9 & 72.7 & 140.2\\
\hline
{\bf Model} & {\bf Parameters:}  & & & &\\
\hline
& $\log L/L_{\odot}$  & & & 2.07 &\\
& $\log T_{\rm eff}$ (K) & & & 5.02 &\\
& $\log g$  & & & 7.0 &\\
& $\log P/k$ (cm$^{-3}$K) & & & 5.85 &\\
\hline
& $V_s$ (km~s$^{-1}$) & & & & 25 \\
& $\log n_{\rm H}$ (cm$^{-3}$) & & & & 2.0 \\
\hline
\hline
\end{tabular}
\end{table*}

\subsection{Reverse Shock Emission}
The photoionisation model presented above fails to correctly predict the strength of both the [O~I] and [N~I] emission lines. Both are under-predicted by a factor 4 or so. However, in Section \ref{Interaction} we discussed the evidence that the PNe is interacting with its local ISM as a result of its motion through this medium. The prime effect of such an interaction is to drive a reverse shock through the outer layers of the nebular shell. It may also drive a forward shock through the ISM, but only if the sound speed in this medium is less than the velocity of the interaction. Otherwise, there is simply a sub-sonic pressure wave in the ISM, which will not be directly observable. Because the shell of the PNe is rather dense, the velocity of the reverse shock will be low; and of order of the expansion velocity. Several 2D and 3D hydrodynamic simulations were constructed to investigate how the interaction between the PNe and ISM affect the outer PN structures, see e.g. Villaver et al. (2002, 2003); M\"{u}ller et al. (2004); Wareing et al. (2007). Here we investigate the emission properties of such a slow shock, and test whether it is capable of producing the observed [O~I] and [N~I] emission.

Since the PNe is appreciably distorted, we assume that the reverse shock velocity is somewhat greater than the expansion velocity. For the purpose of the model we take $V_s = 25$km~s$^{-1}$ and $\log n_{\rm H} = 2.0$ cm$^{-3}$, appropriate to the region of cool un-ionised gas ahead of the ionisation front in the nebula. The implied ram pressure of the ISM is $ P/k \sim1.2\times 10^7$ cm$^{-3}$K.

A slow shock such as this is characterised by a low fractional ionisation of hydrogen at all places through the cooling zone. However, as a consequence of the elevated post-shock temperature ($\sim 21000$K), we have a large collisional contribution to the Balmer decrement. In addition, species such as [O~I] and [N~I], for which the collision strength for excitation increases with temperature, are strongly enhanced. The resultant spectrum predicted by the shock model is shown in Table 4 (column 6). Note that the more highly ionised species are entirely absent in the predicted spectrum, which looks somewhat like an extreme low-excitation Herbig-Haro object.

We find that with only a $\sim 13$\% mixing of shocked gas with the photoionised contribution, we can achieve a much better fit of the de-reddened observational line intensities (Table 4, column 3) with theory (Table 4, column 4).
The shock model provides a surface brightness  in H$\beta$ of $\log S_{\rm H \beta} = -5.41$~erg~cm$^{-2}$. Since the reverse shock can exist only on the leading half of the PNe, we can assume that the total area of the shock is $2 \pi R^2$, where $R$ is the nebular radius. On this basis, we estimate the total shock luminosity $L_{\rm H\beta}= 2.4 \times10^{32}$~erg~s$^{-1}$. This represents 6\% of the total luminosity of PN G342.0-01.7 compared with our empirical shock mixing fraction (13\%) estimated above. Thus we can conclude that the shock model is somewhat consistent with the global energetics of the PN.

\section{Conclusions} \label{Conc}
In this article, we have discussed the results of the first detailed study and optical integral field spectroscopy of the PN G342.0-01.7. The spectra cover both the blue and red spectral ranges and the data is obtained across a field size of 114\arcsec $\times$  75\arcsec. This analysis has provided us with most of the physical and kinematical parameters of the object. We find that the object is a planetary nebula of apparently low excitation class and in an advanced stage of evolution. Plasma diagnostics from both one-zone models and detailed photoionisation modelling revealed an excess in both the nitrogen ($1.0 < N/O <  1.7$) and helium ($0.15 < He/H < 0.16$), placing it in the class of Peimbert Type I PNe, with a fairly massive precursor. Kinematical measurements show that the PNe is expanding at $\sim 20$ km~s$^{-1}$, and is a member of the thin (young) disk population of the Galaxy. A search for the central exciting star was unsuccessful.

We have found clear evidence for an interaction of the expanding nebular shell with the surrounding ISM. The nebula appears to be moving through this ISM in a  south-east direction. The emission line flux, size, physical density and global spectrum of PN G342.0-01.7 are matched by a model with  an effective temperature  $\log T_{\rm eff} \sim 5.05$ and luminosity $1.85 \lesssim \log L \lesssim 2.25$. This lies near the evolutionary tracks of Vassiliadis \& Wood (1993), and is consistent with the inferred dynamical age of $\geq 20000$~yr. The inferred effective temperature of the central star is much higher than would be implied by its excitation class (EC = 0.5). This is found to be a consequence of the evolved state of the nebula and its low ionisation parameter $\log U = -3.4$. Presumably, this mis-match between the temperature estimated from the formal excitation class and the true effective temperature of the central star would also apply to other evolved PNe of this type.

A reverse-shock model designed to account for the interaction of the PNe with its ISM suggests that of order 10\% of the flux from PN G342.0-01.7 is shock-excited. This fraction is consistent with both the global energetics of the PN, and with the observed nebular spectrum - where it is required to explain the anomalous strength of [N~I] $\lambda\lambda 5198,5200$ and [O~I] $\lambda\lambda 6300,6363$ emission lines relative to H$\beta$.

\acknowledgements
The authors would like to thank the anonymous referee for his/her valuable and constructive comments that enabled us to improve the paper. This work was funded by the Deanship of Scientific Research (DSR), King AbdulAziz University, under grant No. (5-130/1433 HiCi). Both Dopita and Vogt wish to thank King Abdulaziz University and the staff of the Astronomy Department during their visits. The DSS-2 images are based on photographic data obtained using The UK Schmidt Telescope. The UK Schmidt Telescope was operated by the Royal Observatory Edinburgh, with funding from the UK Science and Engineering Research Council, until 1988 June, and thereafter by the Anglo-Australian Observatory.  Original plate material is copyright (c) the Royal Observatory Edinburgh and the Anglo-Australian Observatory.  The plates were processed into the present compressed digital form with their permission.  The Digitized Sky Survey was produced at the Space Telescope Science Institute under US Government grant NAG W-2166.

Based on observations made with ESO Telescopes at the La Silla Paranal Observatory under programme ID 089.D-0357(A). This research has made use of NASA's Astrophysics Data System, of \textsc{matplotlib} Hunter (2007), of \textsc{astropy}, a community-developed core \textsc{python} package for Astronomy Astropy (2013), of \textsc{aplpy}, an open-source plotting package for \textsc{python} hosted at {http://aplpy.github.com}, and of \textsc{montage}, funded by the National Aeronautics and Space Administration’s Earth Science Technology Office, Computation Technologies Project, under Cooperative Agreement Number NCC5-626 between NASA and the California Institute of Technology. \textsc{montage} is maintained by the NASA/IPAC Infrared Science Archive.

\end{document}